\documentclass[twocolumn,preprintnumbers,amsmath,amssymb]{revtex4}

\usepackage{graphicx}
\usepackage{dcolumn}
\usepackage{bm}
\usepackage{epstopdf}

\begin{document}


\title{Achieving 50 femtosecond resolution in MeV ultrafast electron diffraction with a double bend achromat compressor}

\author{Fengfeng Qi$^{1,2}$, Zhuoran Ma$^{1,2}$, Lingrong Zhao$^{1,2}$, Yun Cheng$^{1,2}$, Wenxiang Jiang$^{3}$, Chao Lu$^{1,2}$, Tao Jiang$^{1,2}$, Dong Qian$^{3}$, Zhe Wang$^{1,2}$, Wentao Zhang$^{3}$, Pengfei Zhu$^{1,2}$, Xiao Zou$^{1,2}$, Weishi Wan$^{4}$, Dao Xiang$^{1,2,5,*}$ and Jie Zhang$^{1,2}$}
\affiliation{%
$^1$ Key Laboratory for Laser Plasmas (Ministry of Education), School of Physics and Astronomy, Shanghai Jiao Tong University, Shanghai 200240, China \\
$^2$ Collaborative Innovation Center of IFSA (CICIFSA), Shanghai Jiao Tong University, Shanghai 200240, China \\
$^3$ Key Laboratory of Artificial Structures and Quantum Control (Ministry of Education), School of Physics and Astronomy, Shanghai Jiao Tong University, Shanghai 200240, China \\
$^4$ School of Physical Science and Technology, ShanghaiTech University, Shanghai 201210, China \\
$^5$ Tsung-Dao Lee Institute, Shanghai Jiao Tong University, Shanghai 200240, China \\
}
\date{\today}

\begin{abstract}
We propose and demonstrate a novel scheme to produce ultrashort and ultrastable MeV electron beam. In this scheme, the electron beam produced in a photocathode radio-frequency (rf) gun first expands under its own Coulomb force with which a positive energy chirp is imprinted in the beam longitudinal phase space. The beam is then sent through a double bend achromat with positive longitudinal dispersion where electrons at the bunch tail with lower energies follow shorter paths and thus catch up with the bunch head, leading to longitudinal bunch compression. We show that with optimized parameter sets, the whole beam path from the electron source to the compression point can be made isochronous such that the time of flight for the electron beam is immune to the fluctuations of rf amplitude. With a laser-driven THz deflector, the bunch length and arrival time jitter for a 20 fC beam after bunch compression are measured to be about 29 fs (FWHM) and 22 fs (FWHM), respectively. Such an ultrashort and ultrastable electron beam allows us to achieve 50 femtosecond (FWHM) resolution in MeV ultrafast electron diffraction where lattice oscillation at 2.6 THz corresponding to Bismuth $A_{1g}$ mode is clearly observed without correcting both the short-term timing jitter and long-term timing drift. Furthermore, oscillating weak diffuse scattering signal related to phonon coupling and decay is also clearly resolved thanks to the improved temporal resolution and increased electron flux. We expect that this technique will have a strong impact in emerging ultrashort electron beam based facilities and applications. 
\end{abstract}

\maketitle

One of the grand challenges in science is to watch atoms in motion with femtosecond time resolution and Angstrom spatial resolution. MeV ultrafast electron diffraction (UED \cite{UED1,UED3,UCLA,THU,OSAKA,SJTU,BNL,SLAC,DESY}) in which the sample is driven to a non-equilibrium state by a laser and then probed by a delayed electron pulse produced in a photocathode radio-frequency (rf) gun has shown great potential in visualization of atomic dynamics related to lattice vibration \cite{LMO-BNL, SA}, charge density wave \cite{TaS2-BNL, NP}, chemical bond forming and breaking \cite{CHD, CF3I}, etc. With the advent of commercially available ultrashort lasers ($\sim$25 fs FWHM), the temporal resolution in MeV UED depends primarily on the pulse width and timing jitter of the electron beam.

    \begin{figure*}[t]
    \includegraphics[width = 0.8\textwidth]{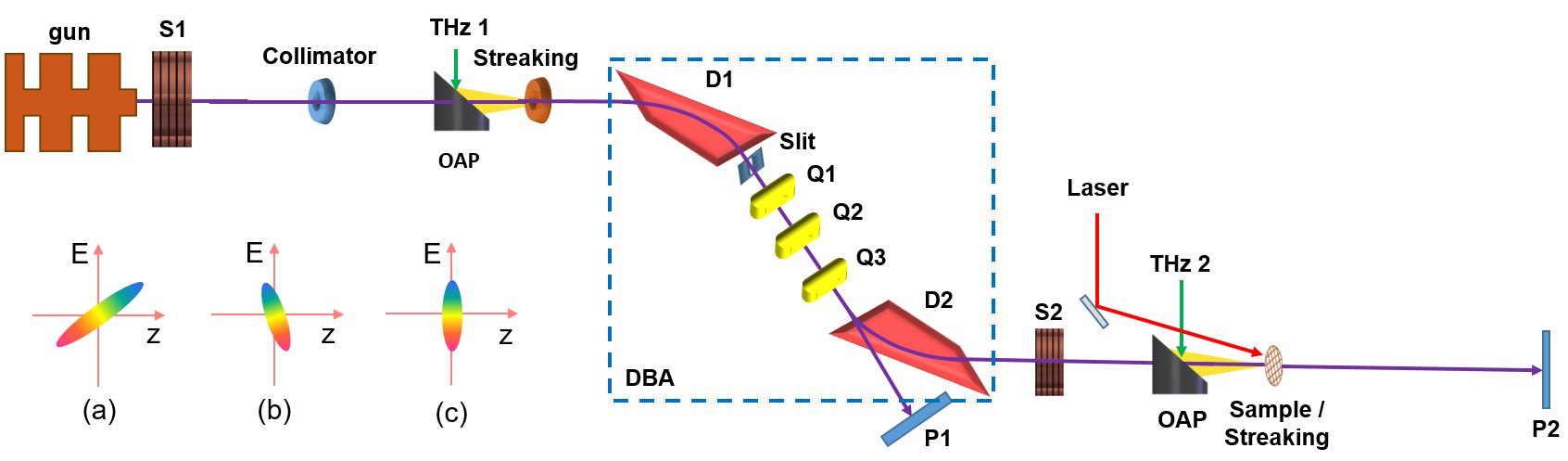}
            \caption{Schematic of DBA based bunch compression. The electron beam produced in a photocathode rf gun is compressed in DBA (region with dashed blue line). Two THz pulses are focused with off-axis parabolic (OAP) mirrors onto circular holes to measure the bunch length and timing jitter. Two solenoid lens (S1 and S2) are used to control the beam size and two phosphor screens (P1 and P2) are used to measure the beam transverse distribution. Electron beam longitudinal phase space evolution from the DBA entrance to the DBA exit, and finally to the sample are illustrated in (a), (b),(c), respectively.    
    \label{Fig.1}}
    \end{figure*}

Tremendous efforts have been devoted in improving the temporal resoution of MeV UED in the past few years, albeit with limited success or application range. For instance, rf buncher has been used to compress MeV beam to below 20 fs (FWHM), but at the cost of increasing the timing jitter to above 100 fs due to rf phase jitter \cite{CUCLA, PRX}. While THz-based time-stamping technique may be used to measure and correct the timing jitter to improve  temporal resolution, the shot-to-shot correction technique only applies to cases where diffraction pattern can be obtained in a single shot \cite{PRX, SLACstreaking, PRLTHz}. Also the detector response time limits the jitter correction technique to low repetition rate, incompatible with the electron hungry experiments that require high repetition rate and long data acquisition time (see, e.g. \cite{CHD, CF3I}). Therefore, it is highly desired if the beam can be compressed with negligible jitter such that time-stamping technique is not needed.

Recently, it has been shown that replacing the rf buncher with a laser-driven THz buncher can be used to compress MeV beams without increasing timing jitter \cite{SJTUC, SLACC}, taking advantage of the fact that the THz pulse is tightly synchronized with the laser and thus has negligible phase error. However, the rf amplitude jitter still limits the timing jitter of THz buncher scheme to $\Delta t=R_{56}\delta E/E$, where $\delta E/E$ is the relative energy stability of the beam at the gun exit and $R_{56}$ is the longitudinal dispersion (a coefficient characterizing how an energy change leads to variation in time of flight) from the buncher to the sample. As a result, the beam timing jitter after THz compression was found to be about 70 fs FWHM in \cite{SJTUC, SLACC}. 

Limited by the rf phase and amplitude jitter, currently the state-of-the-art temporal resolution of MeV UED in integration mode is about 150 fs (FWHM) \cite{CHD, CF3I}, insufficient to study dynamics on sub-100 fs time scale such as Jahn-Teller stretching mode in manganite \cite{LMO}. So far no technique has hitherto successfully compressed a MeV beam to $<$30 fs (FWHM) while simultaneously keeping the timing jitter at a similar level. In this Letter, we demonstrate a novel compression scheme where the electron beam arrival time is immune to both rf phase and amplitude jitter. The technique allows us to compress the electron beam to about 29 fs (FWHM) while keeping the timing jitter at 22 fs (FWHM), achieving 50 fs (FWHM) resolution in MeV UED.  

The experiment is schematically shown in Fig.~1. The electron beam with 3 MeV kinetic energy is produced in a 2.4 cell  photocathode rf (2856 MHz) gun by illuminating a $\sim$100 fs (FWHM) UV laser with transverse size of $\sim$200 $\mu$m (FWHM) on the cathode. The beam first expands under its own Coulomb force with which a positive energy chirp $h=d\delta/dz$ (bunch head having higher energy than bunch tail) is imprinted in the beam longitudinal phase space (Fig.~1a), where $\delta$ is the relative energy difference of an electron with respect to the reference electron. With the bunch head moving faster than the bunch tail, the electron bunch length is increased as it propagates in the drift with negative longitudinal dispersion $R_{56}=c\times dt/d\delta\approx -L/\gamma^2$, where $L$ is the length of the drift and $\gamma$ is the Lorentz factor of the electron. The beam then passes through a double bend achromat (DBA) which consists of two dipole magnets (D1 and D2) and three quadrupole magnets (Q1, Q2, and Q3). Contrary to a drift, the DBA has positive longitudinal dispersion for which electrons with higher energies have longer paths and thus need longer time to pass through. Such a property allows the beam with positive energy chirp to be compressed in a DBA. In our scheme, the beam is over-compressed at the DBA exit where the bunch head and bunch tail switches longitudinal position. As a result, the beam longitudinal phase space rotates counter-clockwise and now has a negative energy chirp (Fig.~1b). Finally, the beam passes through a second drift with again negative longitudinal dispersion. Accordingly, the longitudinal phase space rotates clockwise and the bunch tail exactly catches up with the bunch head (Fig.~1c) at the sample, leading to full compression. The second drift is necessary for UED applications because some space is needed to accommodate the laser injection mirror and sample chamber. 

Assuming the longitudinal dispersion before the DBA entrance to be $R_{56}^b$, that in the DBA to be $R_{56}^d$, and that after the DBA to be $R_{56}^a$, the whole beamline from the gun to the sample can be made isochronous if $R_{56}^b+R_{56}^d+R_{56}^a$=0. In this case the time of flight of the beam is independent of its energy and thus rf amplitudie jitter of the gun will not convert into arrival time jitter. At the same time, if the beam energy chirp from space charge force is controlled in a way such that $1-h(R_{56}^d+R_{56}^a)$=0 is also satisfied, the beam will be compressed at the sample. Note, using rf field to produce energy chirp has the drawback that the rf phase jitter leads to variation of the beam centroid energy, which is further translated to variation of time-of-flight after passing through a section with longitudinal dispersion. With a similar way to produce energy chirp for compression, such timing jitter exists for free-electron lasers as well and timing jitter correction is typically required for achieving a temporal resolution beyond 100 fs (FWHM) \cite{TS2, Coffee1}. In contrast, the positive energy chirp here is produced by space charge force, so the beam centroid energy remains the same in the chirp generation process as dictated by conservation of energy, making this scheme immune also to rf phase jitter. 

In our experiment, the bending angles of the two dipole magnets are 60 degrees and -60 degrees, respectively. The effective length of the bending magnet is about 28 cm. The beam trajectory offset before and after the DBA is 70 cm. With the three quadrupoles set to zero out the transverse dispersion at the DBA exit, we have $R_{56}^d=7.7~$cm for a 3 MeV beam. It should be pointed out that positive longitudinal dispersion can also be obtained with a DBA that bends the beam in the same direction \cite{Korea} or with four-dipole chicane integrated with quadrupoles \cite{DBABNL}. In our experiment, the whole beam line is made isochronous for a 3 MeV beam with $R_{56}^b=-6.0~$cm and $R_{56}^a=-1.7~$cm. 

    \begin{figure}[b]
    \includegraphics[width = 0.48\textwidth]{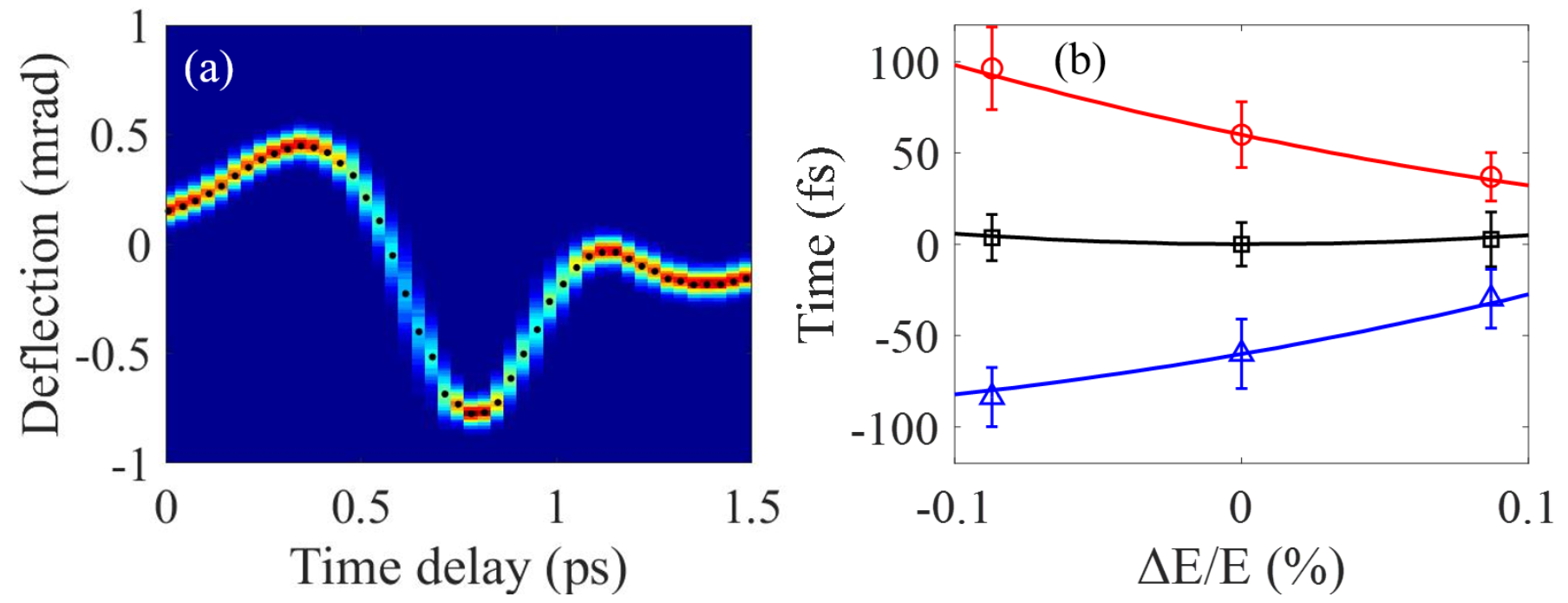}
            \caption{(a) Measured streaking deflectogram as a function of time delay (in 33 fs steps) between the electron beam and THz pulse. (b) Relative beam arrival time measurements (offsets for display) when beam centroid energy is set at 2.8 MeV (red), 3.2 MeV (blue) and 3.0 MeV (black). In this measurement the beam charge is kept low enough such that compression effect is negligible.  
    \label{Fig.2}}
    \end{figure}

To directly measure the beam arrival time at the sample, a vertically polarized single-cycle THz pulse produced by the 800 nm laser in a LiNbO$_3$ crystal \cite{TPFP} is focused to a $100$ microns circular hole to streak the electron beam, similar to a THz deflecting cavity \cite{PRX, SLACstreaking, PRLTHz}. The THz field gives the electron beam a time-dependent angular kick which allows us to map the electron beam time information at the sample into spatial distribution at screen P2. The THz streaking deflectogram is obtained by measuring the beam transverse distribution at various time delay and shown in Fig.~2a. The maximal streaking ramp (around 0.5 ps region in Fig.~2a) is found to be about 4.5 $\mu$rad/fs and the dynamic range where the ramp is approximately constant is about 250 fs. 

With the fluctuation of the beam centroid divergence measured to be 18 $\mu$rad (FWHM), the accuracy of the arrival time measurement is found to be about 4 fs (FWHM). It should be pointed out that because the longitudinal dispersion depends on beam energy, the arrival time at the sample is immune to rf amplitude fluctuation only for a specific beam energy. When the beam energy is set at 2.8 MeV, DBA under-compensates the longitudinal dispersion of the drifts and thus the whole beam line has negative longitudinal dispersion such that higher energy electrons have shorter time of flight. This is shown with red circles in Fig.~2b where the beam energy is varied in the range of $\pm0.1\%$ with its arrival time measured. For 3.2 MeV beam, DBA over-compensates the longitudinal dispersion of the drifts and as a result higher energy electrons have longer time of flight, as shown with blue triangles in Fig.~2b. When the beam energy is set at the designed value (3 MeV), the longitudinal dispersion of the whole beam line is zero and the beam arrival time is independent of the beam energy (black squares in Fig.~2b). The slight nonlinear dependence of the beam arrival time on beam energy in Fig.~2b is due to second order longitudinal dispersion of the DBA, and the experimental results are in excellent agreement with simulations (solid lines in Fig.3b). 

    \begin{figure}[b]
    \includegraphics[width = 0.48\textwidth]{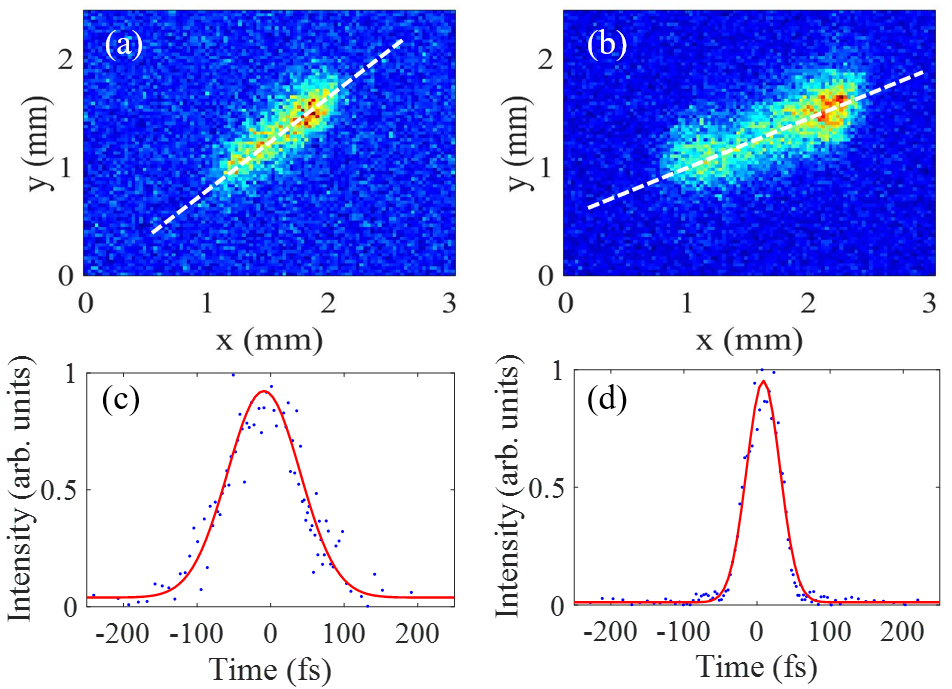}
            \caption{Raw beam distribution measured at screen P1 when the beam energy chirp is low (a) and moderate (b) by controlling the transverse beam size. The bunch length after compression for the low and moderate energy chirp is about 100 fs FWHM (c) and 50 fs FWHM (d), respectively.   
    \label{Fig.3}}
    \end{figure}  

The compression factor, i.e. the ratio of the initial bunch length over the final bunch length, can be written as $1/(1-hR_{56})$. It can be varied by changing the imprinted energy chirp through tuning the beam charge and beam transverse size in the drift section. Because longitudinal space charge force depends on both the transverse position and longitudinal position of the electrons (see, e.g. \cite{ZRH, Subfs, CD}), the energy chirp has both spread and nonlinearity which limit the shortest bunch that can be obtained. In this experiment a collimator before the DBA is used to reduce the spread of the energy chirp by removing the electrons with large transverse offsets. To measure the beam energy chirp at the DBA entrance, the beam is streaked by a second THz pulse (THz 1 in Fig.~1) and then measured at screen P1 with the second bending magnet (D2 in Fig.~1) off. Because the electron beam is streaked vertically and bent horizontally, the vertical axis on P1 becomes the time axis ($y \propto z$) and the horizontal axis becomes the energy axis ($x \propto E$). Two representative beam profiles at P1 for small and moderate energy chirp are shown in Fig.~3a and Fig.~3b, respectively. With the electron bunch length longer than the dynamic range of THz deflector, unambiguous determination of the initial bunch length is difficult. Nevertheless, it can be clearly seen that the energy chirp (i.e. inversely proportional to the slope of the longitudinal phase space which is marked with white dashed line) in Fig.~3b is larger than that in Fig.~3a. Then the second bending magnet of the DBA is turned on to compress the beam and the corresponding beam temporal profiles measured with the THz pulse at the sample are shown in Fig.~3c and Fig.~3d, respectively. With the focusing strength of the solenoid increased, the energy chirp is also increased (Fig.~3b) that together with a larger compression factor leads to a shorter beam (Fig.~3d) after compression even though the initial bunch length is actually longer due to stronger Coulomb repulsion force. 

      \begin{figure}[b]
    \includegraphics[width = 0.48\textwidth]{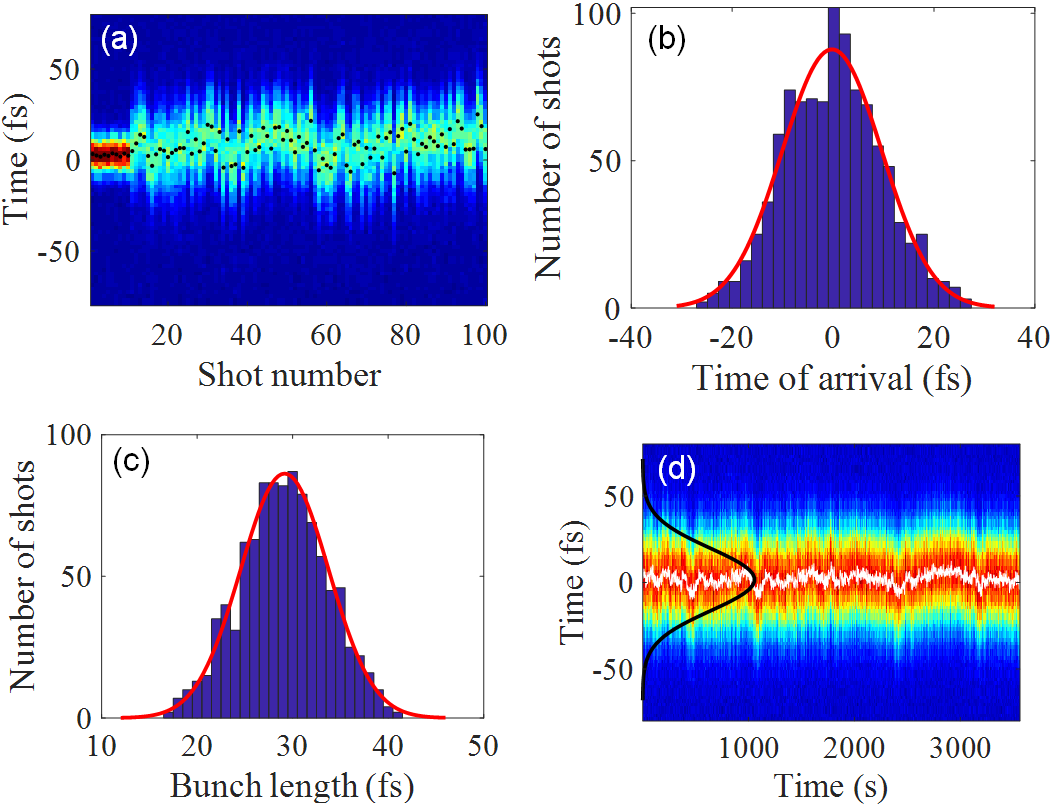}
            \caption{Full compression with DBA. (a) 100 consecutive measurements of the beam temporal profile with THz deflector off (the first 10 shots) and on (the rest 90 shots). Statistics of the beam arrival time (b) and electron beam FWHM pulse width (c) collected over 1000 consecutive shots. Long term stability of the bunch length and arrival time (d). 
    \label{Fig.4}}
    \end{figure}  

To produce an ultrashort electron beam, a slit in the DBA is further used to remove the electrons at bunch head and tail where the energy chirp is dominated by nonlinear curvature \cite{DBAtheory}. After these two collimations, the beam charge is reduced from about 200 fC at the gun exit to about 20 fC after DBA. With the solenoid strength varied to provide the optimal energy chirp for full compression, the electron beam temporal profiles for 100 consecutive shots are measured and shown in Fig.~4a where one can see that the THz deflector has sufficient resolution to resolve the beam time structure. The beam centroid for each individual shot (black dots) is further used to determine the time jitter. The beam timing jitter and bunch length averaged over 1000 consecutive shots are found to be 22 fs (FWHM) and 29 fs (FWHM), respectively. In contrast, the timing jitter at the DBA entrance is estimated to be about 200 fs (FWHM), using the measured energy stability and the longitudinal dispersion. Analysis shows that the residual timing jitter is from the rf phase jitter that leads to variation of beam time-of-flight in the gun. This term can not be compensated for with DBA, but can be eliminated with proper design of the rf gun \cite{DBAtheory}. The bunch length is limited by nonlinearity of the chirp, slice energy spread, as well as high order longitudinal dispersion of the DBA which may be reduced with sextupole magnets. So our results should be considered as representative and further improvements should be feasible with more careful optimization. We also measured the long-term stability of the bunch length and timing jitter, as shown in Fig.~4d. Here each measurement is integrated over 2 seconds, so the distribution in Fig.~4d represents convolution of bunch length and timing jitter. The beam temporal distribution averaged over 1 hour is fitted with a Gaussion distribution (black line in Fig.~4d) with FWHM of 40 fs. This ultrashort and ultrastable beam together with a $\sim$25 fs pump laser allows us to achieve 50 fs (FWHM) resolution in MeV UED.

      \begin{figure}[b]
    \includegraphics[width = 0.48\textwidth]{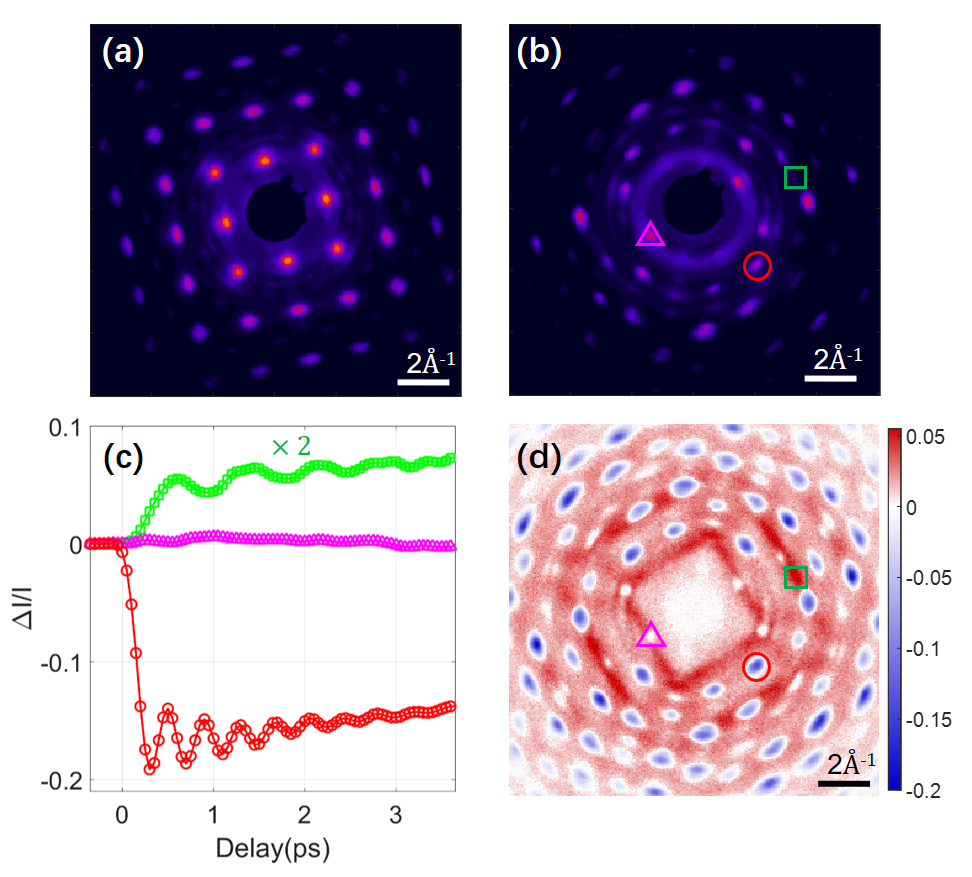}
            \caption{Electron diffraction pattern of Bismuth measured with integration over 100 pulses at normal incidence (a) and with the sample tilted by 26 degrees (b); Diffraction intensity change (c) for the highlighted areas (square, triangle, circle) in Fig.~5b and Fig.~5d, and difference of the diffraction distribution (d) at 0.5 ps time delay between the laser pulse and electron bunch. 
    \label{Fig.5}}
    \end{figure}

The applicability of this scheme was demonstrated in a pump-probe experiment where the compressed beam was used to visualize the phonon dynamics of Bismuth excited by a 800 nm laser with a fluence of about 2.4 mJ/cm$^2$. High quality diffraction pattern measured at normal incidence is shown in Fig.~5a, which demonstrates that the compression process does not lead to considerable degradation to beam emittance and beam pointing jitter. With laser excitation, coherent $A_{1g}$ optical phonons, corresponding to atomic vibrations parallel to the trigonal axis of the rhombohedral unit cell of Bismuth are produced. To see the diffraction spots related to $A_{1g}$ mode, the sample is tilted by about 26 degrees and the corresponding diffraction is shown in Fig.~5b. Diffraction intensities for two Bragg reflections, one sensitive to $A_{1g}$ mode (red circle) and the other not (magenta triangle) are measured and shown in Fig.~5c. The intensity modulation related to $A_{1g}$ mode is clearly seen and the oscillation frequency is found to be about 2.6 THz, consistent with previous results \cite{Science2007, Coffee1}. Furthermore, weak diffuse scattering signals (red stripe-like patterns in Fig.~5d) are also clearly seen which provide momentum-resolved view for electron-phonon coupling and phonon-phonon coupling, complementing the time-resolved inelastic x-ray scattering in FELs \cite{LCLSDS1}. Here, for the green box region (Fig.~5d), the diffuse scattering signal shows oscillation (green line in Fig.~5c) at about 1.3 THz, which may result from anharmonic decay of large-amplitude coherent phonons \cite{LCLSDS2}. Future work on detailed analysis of all the scattering signals may provide more information about how the energy is transferred from the electrons to the lattice and how the phonons interact with each other. These results demonstrate how improvements in temporal resolution and beam brightness may enable new science.

In conclusion, we have demonstrated a novel bunch compression technique that allows us to achieve 50 fs resolution in MeV UED, opening new opportunities for studies of dynamics on sub-100 fs time scale. This technique should be easily transferable to other facilities. In addition to MeV UED, ultrashort and ultrastable electron beam is also essential for external injection in plasma acceleration, THz-driven acceleration, inverse Compton scattering x-ray source, etc. We expect this technique to have a strong impact in future development of ultrashort electron beam based scientific facilities and applications. 

This work was supported by the National Natural Science Foundation of China (Grants No. 11925505, 11504232 and 11674227). One of the authors (DX) would like to thank the support of grant from the office of Science and Technology, Shanghai Municipal Government (No. 16DZ2260200 and 18JC1410700).\\
* dxiang@sjtu.edu.cn

\pagebreak

\end{document}